\documentclass[final]{svjour2}
\usepackage{graphicx}
\usepackage{rotating}
\usepackage{amssymb}
\usepackage{mathptmx}
\usepackage[numbers]{natbib}
\makeatletter
\journalname{Journal of Low Temperature Physics}

\bibpunct{}{}{,}{s}{}{,}

\begin{document}

\newcommand{\ao}{Applied Optics}
\newcommand{\aap}{Astron. and Astroph.}
\newcommand{\mnras}{Mon. Not. R. Astron. Soc.}
\newcommand{\pasp}{Proc. Astron. Soc. of the Pacific}
\newcommand{\jcap}{Journal of Cosmology and Astroparticle Physics}

\newcommand{\hdblarrow}{H\makebox[0.9ex][l]{$\downdownarrows$}-}
\title{WSPEC: A waveguide filter-bank focal plane array spectrometer for millimeter wave astronomy and cosmology}

\author{Sean Bryan$^1$ \and James Aguirre$^2$ \and George Che$^1$ \and Simon Doyle$^3$ \and Daniel Flanigan$^4$ \and Christopher Groppi$^1$ \and Bradley Johnson$^4$ \and Glenn Jones$^4$ \and Philip Mauskopf$^1$ \and Heather McCarrick$^4$ \and Alessandro Monfardini$^5$ \and Tony Mroczkowski$^6$}

\institute{$^1$School of Earth and Space Exploration, Arizona State University, Tempe, AZ 85287 USA\\
\email{sean.a.bryan@asu.edu}\\
$^2$Department of Physics and Astronomy, University of Pennsylvania, Philadelphia, PA 19104 USA\\
$^3$School of Physics and Astronomy, Cardiff University, Cardiff, CF24 3AA, UK\\
$^4$Department of Physics, Columbia University, New York, NY 10027 USA\\
$^5$Institut N\'{E}EL, Grenoble, France\\
$^6$Naval Research Laboratory, Washington DC 20375 USA}
\maketitle

\begin{abstract}

Imaging and spectroscopy at (sub-)millimeter wavelengths are key frontiers in astronomy and cosmology. Large area spectral surveys with moderate spectral resolution ($R$=50-200) will be used to characterize large scale structure and star formation through intensity mapping surveys in emission lines such as the CO rotational transitions. Such surveys will also be used to study the SZ effect, and will detect the emission lines and continuum spectrum of individual objects. WSPEC is an instrument proposed to target these science goals. It is a channelizing spectrometer realized in rectangular waveguide, fabricated using conventional high-precision metal machining. Each spectrometer is coupled to free space with a machined feed horn, and the devices are tiled into a 2D array to fill the focal plane of the telescope. The detectors will be aluminum Lumped-Element Kinetic Inductance Detectors (LEKIDs). To target the CO lines and SZ effect, we will have bands at 135-175 GHz and 190-250 GHz, each Nyquist-sampled at $R\approx$200 resolution. Here we discuss the instrument concept and design, and successful initial testing of a WR10 (i.e. 90 GHz) prototype spectrometer. We recently tested a WR5 (180 GHz) prototype to verify that the concept works at higher frequencies, and also designed a resonant backshort structure that may further increase the optical efficiency. We are making progress towards integrating a spectrometer with a LEKID array and deploying a prototype device to a telescope for first light.


\end{abstract}

\section{Introduction}

Large area millimeter wave spectral surveys with moderate spectral resolution can be used to characterize the large scale structure and star formation history of the universe using intensity mapping of emission lines such as CO \cite{lidz11} and CII \cite{silva14}. Multi-object mm-wave wide-band spectroscopy with moderate spectral resolution will enable galaxy redshift surveys. Further studies of hot gas in galaxy clusters through the Sunyaev Zeldovich (SZ) effect will be enabled by high angular resolution and moderate spectral resolution instruments.

Achieving these science goals requires a millimeter wave instrument with moderate resolving power, single-dish-class angular resolution, and high instantaneous sensitivity. Z-spec \cite{zspec}, an $R\sim100$ single-pixel waveguide grating spectrometer, has demonstrated photon noise limited performance in the atmospheric window from 200-300 GHz. Many programs, such as SuperSpec \cite{HaileyDunsheath}, Micro-Spec \cite{Cataldo}, DESHIMA \cite{Endo}, and TIME \cite{staniszewski14}, are currently working towards on-chip spectrometer technology to enable focal plane arrays. In this paper, we describe a waveguide filter bank spectrometer, WSPEC, which can be fabricated with existing high-precision machining techniques and is an alternative technology for enabling filled focal plane array spectrometers. The detectors are single-layer aluminum LEKIDs \cite{mccarrick+2014} similar to those described by Heather McCarrick and Daniel Flanigan elsewhere in these proceedings. Aluminum LEKIDs have successfully been used on the IRAM telescope for science observations with the NIKA instrument\cite{mauskopf14}, and have also been sucessfully built and demonstrated for the NIKA2 instrument as described by Johannes Goupy elsewhere in these proceedings.

\begin{figure}
\begin{center}
\includegraphics[width=0.9\linewidth]{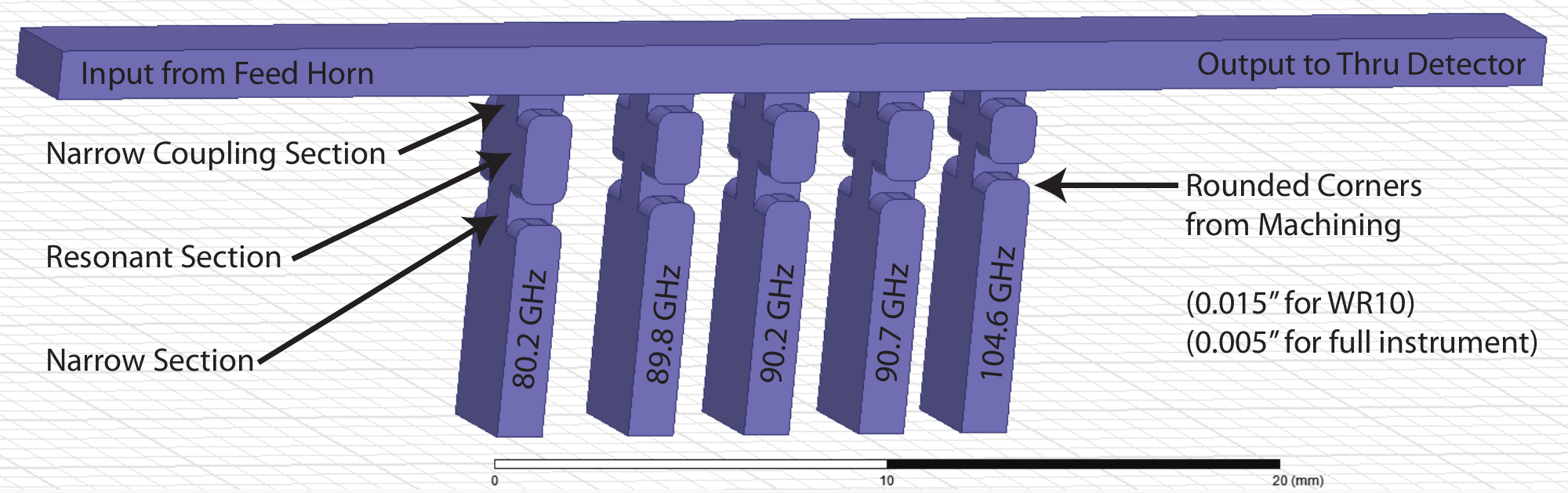}
\end{center}
\caption{Illustration of the spectrometer concept. Light from a feed horn (not shown) couples into the main waveguide, and different $R\approx$200 frequency bands are selected off by each of the five channels shown.}
\label{prototype_overview}
\end{figure}

The spectrometer concept is shown in Figure~\ref{prototype_overview}, which is a drawing of a prototype WR10 device we built and tested\cite{bryan15}. Broadband light from the sky is coupled into a main waveguide using a feed horn. At an individual spectrometer channel, there is a narrow coupling section of waveguide, which is narrow enough that its cutoff frequency is 50\% higher than the device operating frequency. This is followed by a half-wave resonant section of full-width waveguide, and another narrow section to define the end of the resonator. On resonance, the evanescent coupling through the narrow sections is effectively capacitive, and the resonance in the half-wave section is effectively inductive, which means that overall the channel is matched with the main waveguide and power passes through to that channel's detector. Off-resonance, this cancellation does not take place, which keeps out of band radiation from coupling to the detector. The design of a full 54-channel device that fills the single-moded passband of rectangular waveguide is discussed in an earlier publication\cite{bryan15}.

\section{Including Detector Effects in the Model}

The design and performance modeling of the spectrometer channels was performed with HFSS. This yields a three-port S-matrix model of each spectrometer channel, which we cascaded together into a complete spectrometer model \cite{bryan15}. Since it is purely a S-matrix model, this yielded a performance estimate that assumes perfect terminations at all ports of the device. However, even with high optical-efficiency LEKID detectors, the termination at the detector ports of the real device will not be perfect.

To model this imperfect termination effect in the integrated 54-channel device, and verify that it will not significantly impact performance, we took the three-port individual-channel HFSS models, and terminated the detector port in each device with a HFSS simulation of the input impedance of a LEKID detector. This turned the three-port model of a single channel into a two-port model with some loss due to the power absorbed by the detector. In addition, this two-port model also includes the effect of any small reflections caused by imperfect absorption by the LEKID detector.

\begin{figure}
\begin{center}
\includegraphics[width=0.525\linewidth]{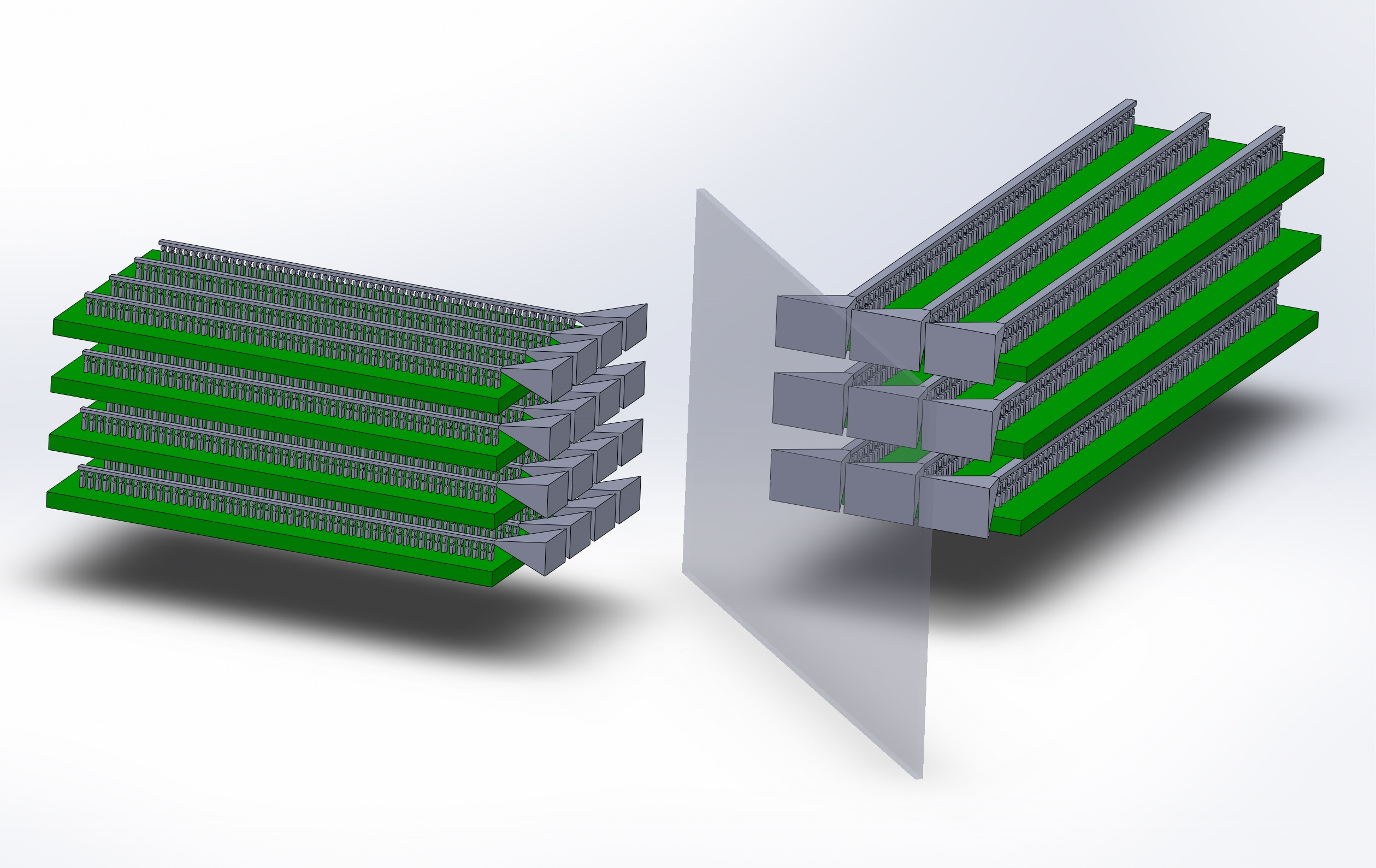}
\includegraphics[width=0.375\linewidth]{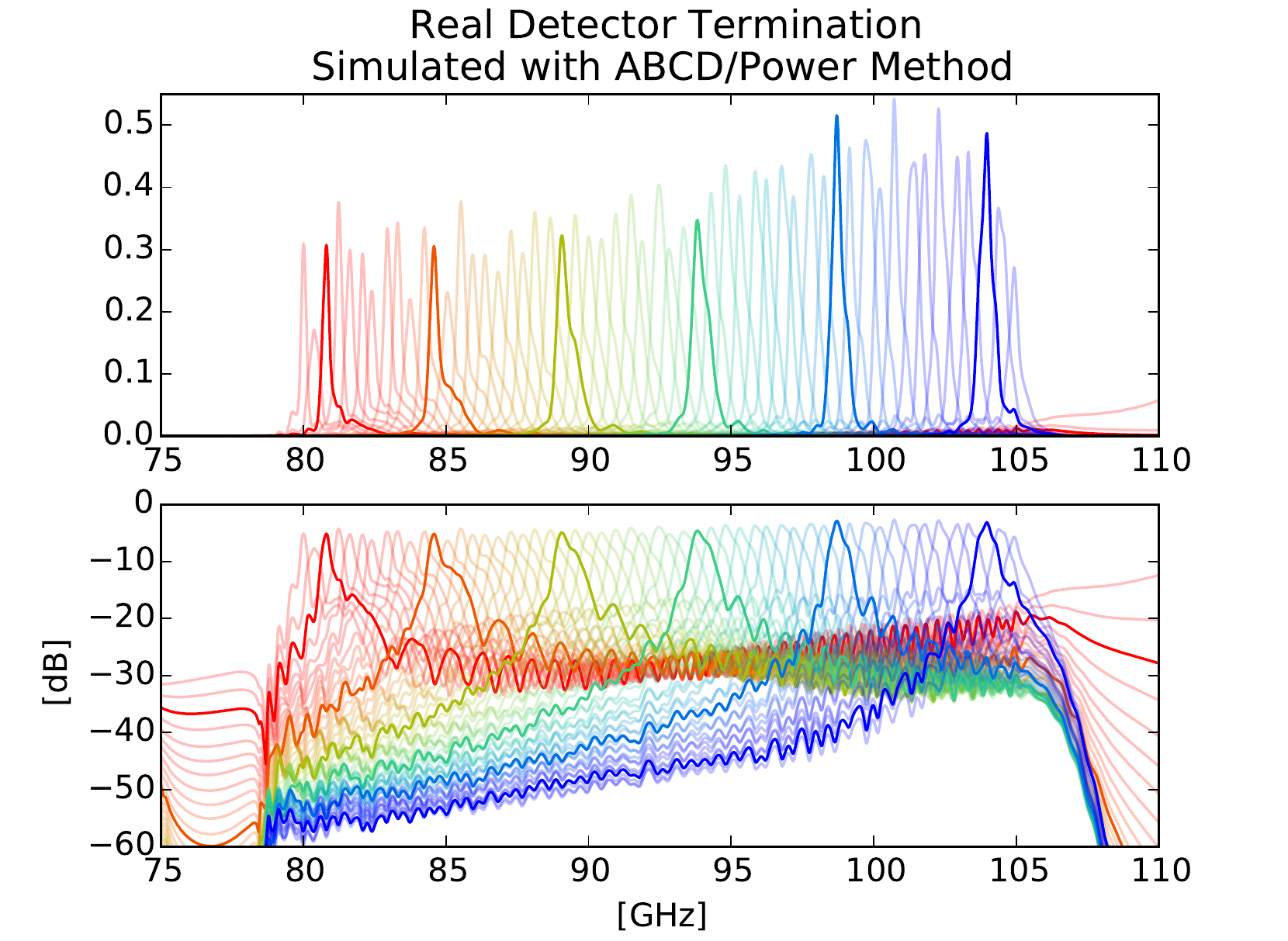}
\end{center}
\caption{Simulated passbands (right) of a full 54-channel device, including the effect of imperfect detector terminations. (The WR10 simulation is shown, but in the low-loss limit only the frequency axis needs scaling to model a device at another band.) The drawing to the left shows an illustration of a two-band WR4 / WR6 focal plane array concept using a dichroic to split the bands. A negative of each individual spectrometer is shown in grey, approximately three inches in total length. The LEKID detector cards are illustrated in green.}
\label{full_spectrometer_simulation_and_fpu}
\end{figure}

We developed an approach to track the power absorbed at each individual detector in the integrated device. Consider the $i$th detector in the model. We cascaded the individual spectrometer channels above that detector into a combined transmission line matrix $ABCD_{subsequent}$, and cascaded the others, from the first to the $i$th, into a matrix $ABCD_{before}$. We then converted the transmission line matrix $ABCD_{subsequent}$ into an S-matrix, which yielded the input impedance looking into these channels. Using this impedance at port 2, and the nominal waveguide impedance at port 1, we then converted $ABCD_{before}$ to its equivalent S-matrix. Absorption in a detector $a_i$, transmission on to the subsequent detectors $t\equiv S_{21}\times\mathrm{conj}(S_{21})$, or reflection from the entire structure $r\equiv S_{11}\times\mathrm{conj}(S_{11})$ are the only things that can happen in the first $i$ detectors. So, we simply subtracted the transmission and reflection to leave the absorption, that is, $a_i = 1 - r - t$.

This quantity $a_i$ is actually the combined absorption of all the detectors from the first to the $i$th. To get a model for an individual detector's absorbed power, we repeated the calculation for the detectors from the first to the $(i-1)$th to calculate $a_{i-1}$. The power absorbed at the $i$th detector alone is therefore $a_i - a_{i-1}$. We then repeated this calculation for each channel in the spectrometer, as well as sweeping the simulation frequency, to yield a complete model of the integrated spectrometer system and its LEKID detectors. The results are shown in Figure~\ref{full_spectrometer_simulation_and_fpu}. In the end, the overall optical efficiency of the device remains high, and the effect of imperfect detector terminations is not calculated to induce significant out of band coupling or other problems in the spectrometer. This modeling motivates moving forward with building a device with integrated LEKIDs.

\section{WR5 Prototype Test Results}

\begin{figure}
\begin{center}
\includegraphics[width=0.45\linewidth]{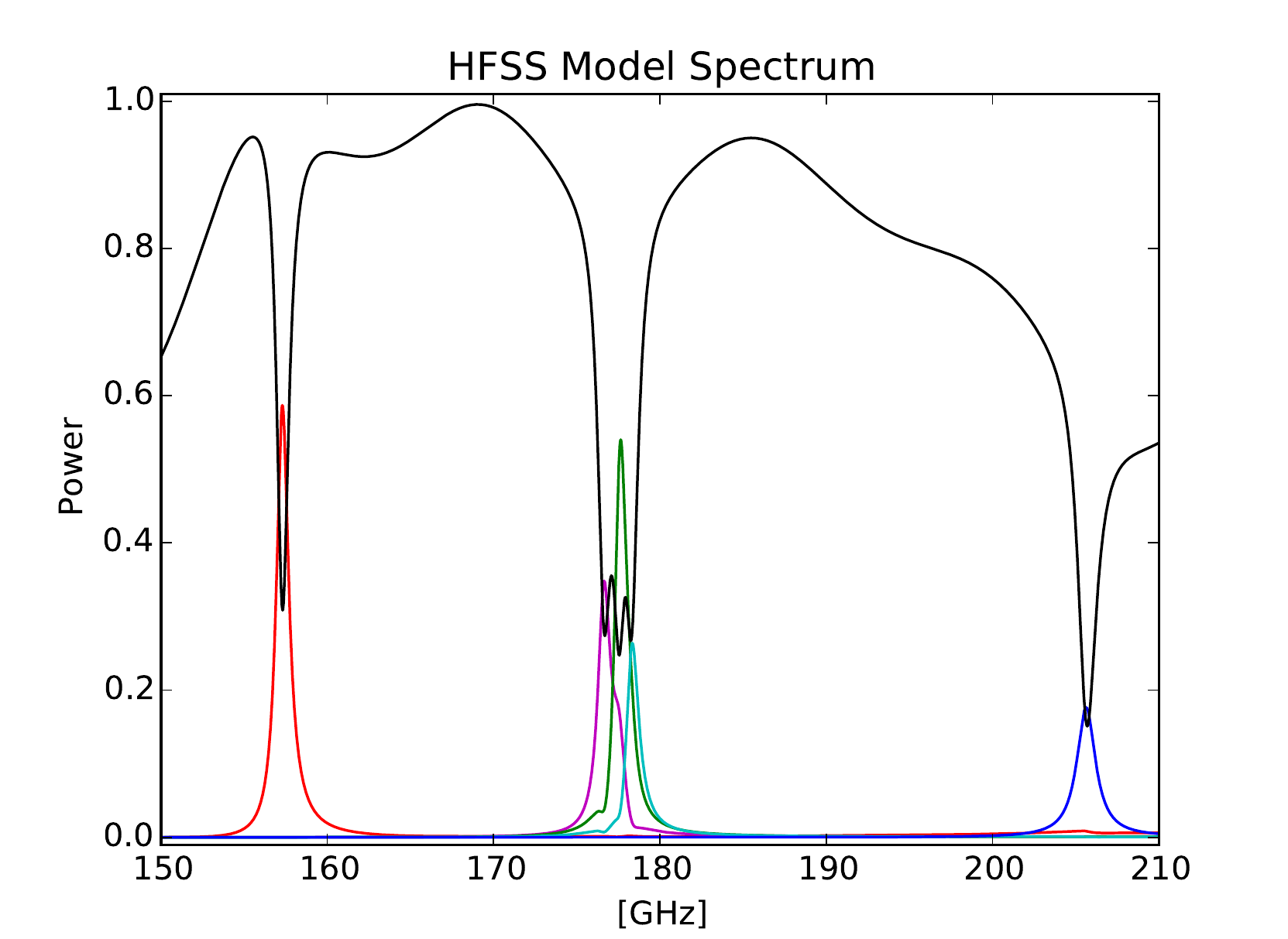}
\includegraphics[width=0.45\linewidth]{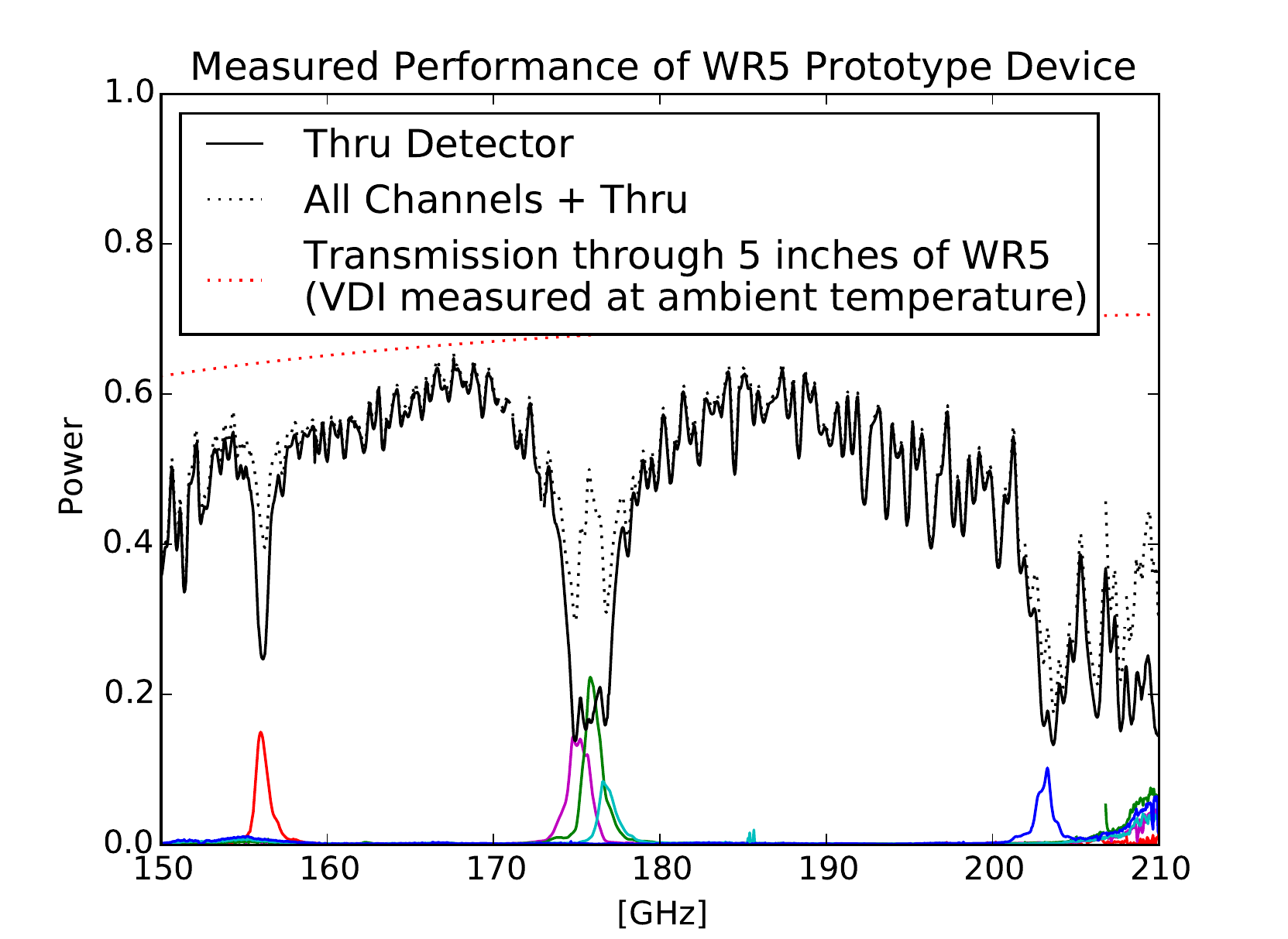}

\end{center}
\caption{Lossless model (left) and measured (right) performance of the WR5 prototype. The two broadly agree, except for an observed decrease in optical efficiency that is consistent with the known impact of waveguide loss. This will be reduced in the science device due to its smaller size, and we may coat the device in superconducting niobium to eliminate waveguide loss in our band. The source power is very low above about 205 GHz, so the apparent increase in coupling in the measured data is likely a measurement artifact.}
\label{wr5_test}
\end{figure}

\begin{table}
\begin{center}
\begin{tabular}{lllllll}
Measured   & Model & Meas. & Model & Meas. & Scaled & Model \\
Frequency  & Frequency  & $R$     & $R$     & OE    & OE     & OE     \\ \hline \\
156.00 GHz & 157.30 GHz & 186   & 182   & 15$\%$    & 58$\%$     & 59$\%$     \\
175.23 GHz & 176.65 GHz & 119   & 120   & 14$\%$    & 31$\%$     & 35$\%$     \\
175.81 GHz & 177.65 GHz & 156   & 193   & 22$\%$    & 62$\%$     & 54$\%$     \\
176.59 GHz & 178.33 GHz & 161   & 207   & \phantom{1}8$\%$     & 24$\%$     & 26$\%$     \\
203.30 GHz & 205.63 GHz & 220   & 154   & 10$\%$    & 21$\%$     & 18$\%$     \\ \hline
\end{tabular}
\end{center}
\caption{Modeled and calculated performance of the WR5 prototype. The measured center frequencies and spectral resolutions $R$ of each channel agree broadly with the model. The measured and modeled optical efficiencies (OE) are shown, as well as the measured optical efficiency scaled up by a calculated value to remove the impact of waveguide loss. This value includes the physical length (5 inches) of the waveguides between the flanges and the device, as well as the $R \times (\lambda/2)$ effective electrical length of the resonator cavities. Since our science device will be smaller and possibly coated with superconducting niobium, waveguide loss will have a significantly smaller impact on the science device performance.}
\label{wr5_table}
\end{table}

To test that the spectrometer concept will work at the $\sim200$ GHz frequencies that are needed for science observations, we built a five-channel prototype device in WR5 waveguide. The device is nearly identical to the WR10 prototype built initially, but the design is scaled down. To fit a standard UG-387/U flange at each detector port, the entire prototype device needed to be five inches in length, which is significantly longer than the three inches a science device would be. This increased length would significantly increase the impact of any waveguide loss on the prototype system. We measured the performance in the lab with the apparatus described in a previous article\cite{bryan15}. A synthesizer generates a single microwave frequency which drives the source port of a WR5 VNA extender. The spectrum is measured by sweeping the microwave frequency, and recording the resulting signal at a diode power detector placed at the desired spectrometer output port. The measured performance of the device, shown in Figure~\ref{wr5_test} and Table~\ref{wr5_table}, is broadly consistent with the model, and the overall reduced optical efficiency observed is consistent with the known impact of waveguide loss in the device. This will be reduced in the science device due to its smaller size, and we may sputter a superconducting niobium coating to the device to eliminate waveguide loss in this band altogether.

\section{Individual Channel Backshorts}

Like all filter bank spectrometers, the maximum possible optical efficiency of an individual WSPEC channel is approximately 50$\%$. This is because if both the detector port, and the port that continues to the other spectrometer channels, are perfectly matched (i.e. the best possible case), the power will divide evenly between the two. Creating a filter bank with channels with closer spectral spacing than Nyquist sampling can boost the effective optical efficiency slightly higher. However, drawbacks of this approach include increased channel count and some redundancy between spectral channels.

\begin{figure}
\begin{center}
\includegraphics[width=0.525\linewidth]{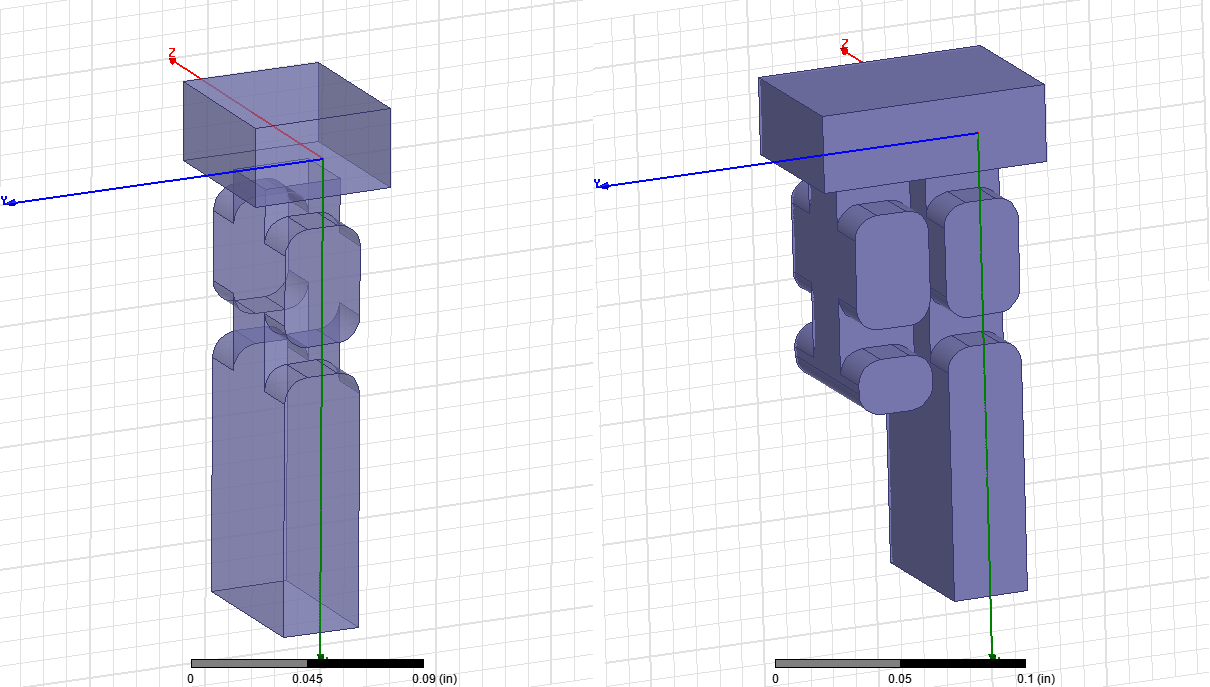}
\includegraphics[width=0.375\linewidth]{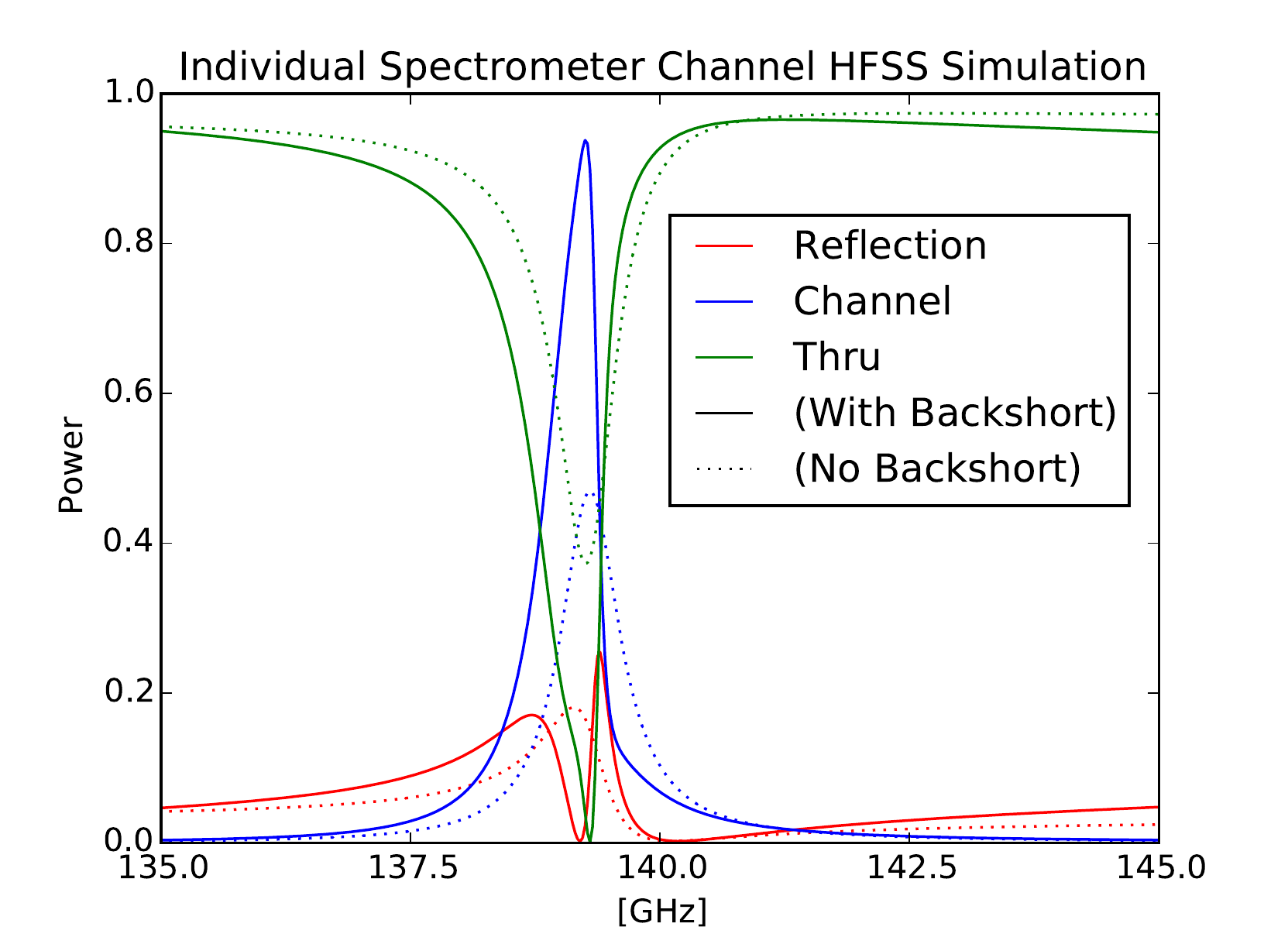}
\end{center}
\caption{Drawing (left) and HFSS simulation (right) of an individual spectrometer channel with a resonant backshort. Adding a second copy of the spectrometer channel further down the main waveguide, but terminating it with a short, reflects nearly all of the in-band power onto the spectrometer channel's detector, well beyond the nominal filter bank limit of 50\%.}
\label{invididual_backshort}
\end{figure}

To go beyond this 50$\%$ limit, we are considering placing a resonant backshort behind each individual channel in the spectrometer. As shown in Figure~\ref{invididual_backshort}, placing an identical copy of the spectrometer channel a quarter-wavelength further down the main transmission line, but terminating this channel in a short instead of an absorber, increases the optical efficiency to nearly unity. Conceptually, this means that on resonance, the power that goes past the detector channel will be reflected from the short in the second channel, and couple back down the original channel to its detector. The distance between the detector channel and the backshort is selected to couple the power down the detector port, which prevents it from coupling back out of the feed horn. The HFSS geometry, and a simulation of an individual channel both with and without an individual backshort, is shown in Figure~\ref{invididual_backshort}.

\section{Conclusions}

Focal plane arrays of millimeter-wave spectrometers will soon enable key measurements of cosmic structure, star formation history, and studies of individual high-redshift objects. Complementing other approaches, the straightforward fabrication of these devices and the ability to make room temperature prototypes has enabled us to make significant progress developing WSPEC. In these proceedings, we presented a design and full model that includes both the spectrometer and the LEKIDs, promising measurements in a WR5 prototype that shows the concept works well at 200 GHz, and a first design for a concept that may further increase the optical efficiency beyond the nominal 50$\%$ limit for a filter bank spectrometer. The next step is building a prototype device with LEKID detectors, which will be a key step towards building a instrument to bring to a telescope for first light.

\begin{acknowledgements}
This work is partially supported by the LabEx FOCUS.
\end{acknowledgements}

\bibliographystyle{unsrt}
\bibliography{bibliography}

\begin{thebibliography}{10}

\bibitem{lidz11}
A.~Lidz et~al.
\newblock Intensity mapping with carbon monoxide emission lines and the
  redshifted 21 cm line.
\newblock {\em Ap. J.}, 741(2):70, 2011.

\bibitem{silva14}
M.~B.~Silva et~al.
\newblock Prospects for detecting cii emission during the epoch of
  reionization.
\newblock {\em ArXiv:1410.4808}, October 2014.

\bibitem{zspec}
H.~Inami et~al.
\newblock {A broadband millimeter-wave spectrometer Z-spec: sensitivity and
  ULIRGs}.
\newblock In {\em SPIE Astronomical Telescopes and Instrumentation}, volume
  7020 of {\em Proc. SPIE}, July 2008.

\bibitem{HaileyDunsheath}
S.~Hailey-Dunsheath et~al.
\newblock {Status of SuperSpec: a broadband, on-chip millimeter-wave
  spectrometer}.
\newblock In {\em SPIE Astronomical Telescopes and Instrumentation}, volume
  9153 of {\em Proc. SPIE}, August 2014.

\bibitem{Cataldo}
G.~Cataldo et~al.
\newblock {Micro-Spec: an ultracompact, high-sensitivity spectrometer for
  far-infrared and submillimeter astronomy}.
\newblock {\em Appl. Opt.}, 53:1094, February 2014.

\bibitem{Endo}
A.~Endo et~al.
\newblock {Development of DESHIMA: a redshift machine based on a
  superconducting on-chip filterbank}.
\newblock In {\em SPIE Astronomical Telescopes and Instrumentation}, volume
  8452 of {\em Proc. SPIE}, September 2012.

\bibitem{staniszewski14}
Z.~Staniszewski et~al.
\newblock {The Tomographic Ionized-Carbon Mapping Experiment (TIME) CII Imaging
  Spectrometer}.
\newblock {\em Low Temp. Phys.}, 176:767--772, September 2014.

\bibitem{mccarrick+2014}
H.~McCarrick et~al.
\newblock Horn-coupled, commercially-fabricated aluminum lumped-element kinetic
  inductance detectors for millimeter wavelengths.
\newblock {\em Rev. Sci. Inst.}, 85(12):--, 2014.

\bibitem{mauskopf14}
P.~D.~Mauskopf et~al.
\newblock Photon-noise limited performance in aluminum lekids.
\newblock {\em Low Temp. Phys.}, 176(3-4):545--552, 2014.

\bibitem{bryan15}
S.~Bryan et~al.
\newblock A compact filter-bank waveguide spectrometer for millimeter
  wavelengths.
\newblock {\em Terahertz Science and Technology, IEEE Transactions on},
  5(4):598--604, July 2015.

\end{thebibliography}

\end{document}